\documentclass[natbib]{svcyclop}

\usepackage{color}
\usepackage{graphicx}    
\usepackage{amsmath,amsfonts}

\newcommand{\R}{\ensuremath{{\mathbb R}}}

\newtheorem{thm}{\bf Theorem}
\newtheorem{assumption}{\bf Assumption}

\begin{document}

\title{Disturbance Observer}

\author{Hyungbo Shim}

\institute{Department of Electrical and Computer Engineering, \\ Seoul National University, Seoul, Korea \\ 
+82-2-880-1745 / {\tt hshim@snu.ac.kr} 
}

\maketitle

\section{Abstract}

Disturbance observer is an inner-loop output-feedback controller whose role is to reject external disturbances and to make the outer-loop baseline controller robust against plant's uncertainties.
Therefore, the closed-loop system with the DOB approximates the nominal closed-loop by the baseline controller and the nominal plant model with no disturbances.
This article presents how the disturbance observer works under what conditions, and how one can design a disturbance observer to guarantee robust stability and to recover the nominal performance not only in the steady-state but also for the transient response under large uncertainty and disturbance.

\section{Keywords}

robust stabilization; robust transient response; disturbance attenuation; singular perturbation; normal form

\section{Introduction}

\begin{figure}[h]
	\centering
	\includegraphics[height=4cm]{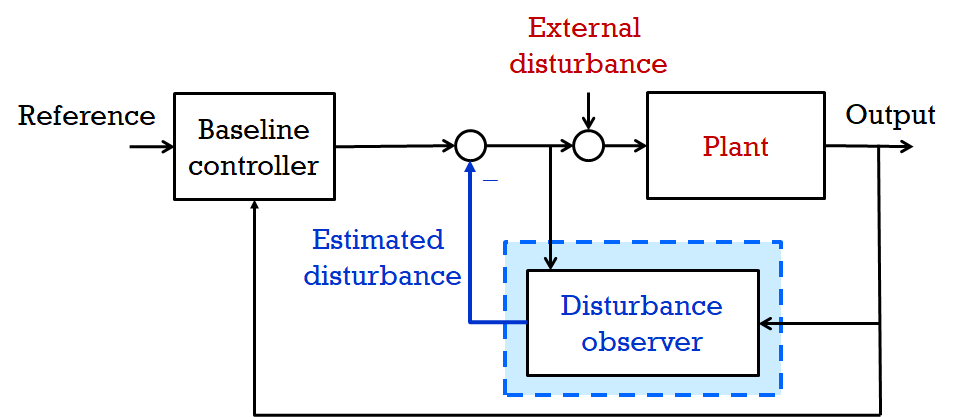}
	\caption{Disturbance observer as an inner-loop feedback controller}
	\label{fig:1}       
\end{figure}

The term {\em disturbance observer} has been used for a few different algorithms and methods in the literature.
In this article, we restrict the disturbance observer to imply the inner-loop controller as described conceptually in Fig.~\ref{fig:1}.
This is one that has been employed in many practical applications and verified in industry, and is often simply called ``DOB.''

%
%

The primary goal of DOB is to estimate the external disturbance at the input stage of the plant, which is then used to counteract the external disturbance as in Fig.~\ref{fig:1}.
This initial idea is extended to deal with the plant's uncertainties when uncertain terms can be lumped into the external disturbance.
Therefore, DOB is considered as a method for robust control.
More specifically, DOB robustifies the (possibly non-robust) baseline controller.
The baseline controller is supposed to be designed for a nominal model of the plant that does not have external disturbances nor uncertainties.
By inserting the DOB in the inner-feedback-loop, the nominal stability and performance that would have been obtained by the baseline controller and the nominal plant, can be approximately recovered even in the presence of disturbances and uncertainties.
There is effectively no restriction on the baseline controller as long as it stabilizes the nominal plant.

When there is no uncertainty and no disturbance with a DOB being equipped, the closed-loop system recovers the nominal performance completely, and as the amount of uncertainty and disturbance grows, the performance degrades gradually while it is still close to the nominal performance.
This is in contrast to other robust controls based on the worst-case design, which sacrifices the nominal performance for the worst uncertainty.

\section{Disturbance Observer for Linear System}

\begin{figure}[h]
	\centering
	\includegraphics[height=4cm]{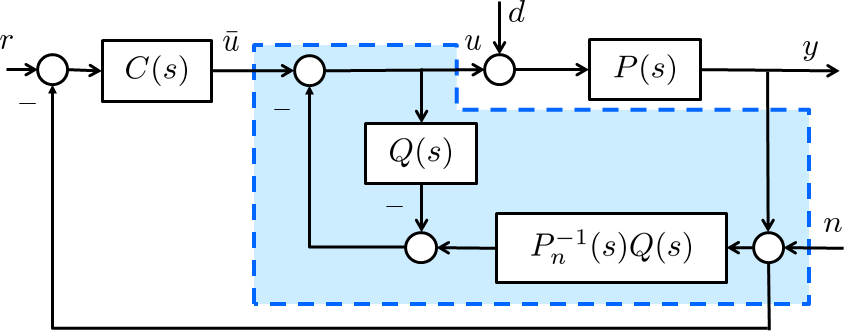}
	\caption{Disturbance observer for a linear plant}
	\label{fig:lindob}       
\end{figure}

Let $P(s)$ be the transfer function of a unknown linear plant, $P_n(s)$ be its nominal model, and $C(s)$ be a baseline controller designed for the nominal model $P_n(s)$.
Then, the closed-loop system with the DOB can be depicted as Fig.~\ref{fig:lindob}.
In the figure, $Q(s)$ is a stable low-pass filter whose dc gain is one.
The relative degree (i.e., the order of denominator minus the order of numerator) of $Q(s)$ is greater than or equal to the relative degree of $P_n(s)$, so that the block $P_n^{-1}(s)Q(s)$ is proper and implementable.
The signals $d$ and $r$ are the external disturbance and the reference, respectively, which are assumed to have little high frequency components, 
and the signal $n$ is the measurement noise. 

From Fig.~\ref{fig:lindob}, the output $y$ in the frequency domain is written as
\begin{multline}\label{eq:ys}
y(s) = \frac{P_n P C}{P_n(1+PC) + Q(P-P_n)} r(s) \\
+ \frac{P_n P (1-Q)}{P_n(1+PC) + Q(P-P_n)} d(s) - \frac{P(Q+P_nC)}{P_n(1+PC) + Q(P-P_n)}n(s).
\end{multline}
Assume that all the transfer functions are stable (i.e., all the poles have negative real parts), and let $\omega_c$ be the cut-off frequency of the low-pass filter so that $Q(j\omega) \approx 1$ for $\omega \ll \omega_c$ and $Q(j\omega) \approx 0$ for $\omega \gg \omega_c$.
Therefore, it is seen from \eqref{eq:ys} that, for $\omega \ll \omega_c$, 
\begin{equation}\label{eq:linrecover}
y(j\omega) \approx \frac{P_nC}{1+P_nC} r(j\omega) - n(j\omega) 
\end{equation}
and for $\omega \gg \omega_c$ where $d(j\omega) \approx 0$ and $r(j\omega) \approx 0$,
$$y(j\omega) \approx \frac{PC}{1+PC} r(j\omega) + \frac{P}{1+PC} d(j\omega) - \frac{PC}{1+PC} n(j\omega) \approx - \frac{PC}{1+PC} n(j\omega).$$
The property \eqref{eq:linrecover} is of particular interest because the input-output relation recovers the nominal closed-loop system without being affected by the disturbance $d$.

The low-pass filter $Q(s)$ is often called `Q-filter' and it is typically given by
$$Q(s) = \frac{a_0}{ (\tau s)^\nu + a_{\nu-1} (\tau s)^{\nu-1} + \cdots + a_1 (\tau s) + a_0 }$$
where $\nu$ is the relative degree of $P_n(s)$, the coefficients $a_i$ are chosen such that $s^{\nu} + a_{\nu-1}s^{\nu-1} + \cdots + a_0$ be a Hurwitz polynomial, and the constant $\tau > 0$ determines the cut-off frequency.

\subsection{Robust stability condition}

The beneficial property \eqref{eq:linrecover} is obtained under the assumption that the transfer functions in \eqref{eq:ys} are stable for all possible uncertain plants $P(s)$.
Therefore, it is necessary to design $Q(s)$ (or, choose $\tau$ and $a_i$'s) such that the closed-loop system remains stable for all possible $P(s)$.

In order to deal with uncertain plants having parametric uncertainty, consider the set ${\mathcal P}$ of uncertain transfer functions
\begin{multline*}
{\mathcal P} = \left\{ P(s) = g \frac{s^{n-\nu} + \beta_{n-\nu-1}s^{n-\nu-1} + \cdots + \beta_0}{s^n + \alpha_{n-1}s^{n-1} + \cdots + \alpha_0} : \alpha_i \in [\underline \alpha_i, \overline \alpha_i], \beta_i \in [\underline \beta_i, \overline \beta_i], g \in [\underline g,\overline g] \right\}
\end{multline*}
where the intervals $[\underline \alpha_i, \overline \alpha_i]$, $[\underline \beta_{i}, \overline \beta_{i}]$, and $[\underline g, \overline g]$ are known, and $\underline g > 0$.

\begin{thm}[\citet{SJ09}]\label{thm:1}
Assume that (a) the nominal model $P_n$ belongs to ${\mathcal P}$ and the baseline controller $C$ internally stabilizes $P_n$, (b) all transfer functions in ${\mathcal P}$ are minimum phase, and (c) the coefficients $a_i$ of $Q$ are chosen such that
$$p_f(s) := s^\nu + a_{\nu-1} s^{\nu-1} + \cdots + a_1 s + \frac{g}{g^*} a_0$$
is Hurwitz for all $g \in [\underline g,\overline g]$, where $g^* \in [\underline g,\overline g]$ is the nominal value of $g$,
then, there exists $\tau^*>0$ such that, for all $\tau \le \tau^*$, the transfer functions in \eqref{eq:ys} are stable for all $P(s) \in {\mathcal P}$.
\end{thm}

The value of $\tau^*$ can be computed from the knowledge of the bounds of the intervals in ${\mathcal P}$ \citep{SJ09}, but it may also be conservatively chosen based on repeated simulations in practice.
Smaller $\tau^*$ implies larger bandwidth of Q-filter, which is desired in the sense that the property \eqref{eq:linrecover} holds for larger frequency range.

The proof of Theorem \ref{thm:1} proceeds by observing the closed-loop system in Fig.~\ref{fig:lindob} has $2n+m$ poles where $m$ is the order of $C(s)$.
Then one can inspect the behavior of those poles by changing the design parameter $\tau$.
For this, let us denote the poles by $\lambda_i(\tau)$, $i=1,\dots,2n+m$.
Then, it can be shown that
\begin{equation}\label{eq:lambda}
\lim_{\tau \to 0} \tau \lambda_i(\tau) = \lambda_i^*, \quad i = 1, \dots, \nu
\end{equation}
where $\lambda_i^*$, $i=1,\dots,\nu$, are the roots of $p_f(s)$, and
$$\lim_{\tau \to 0} \lambda_i(\tau) = \lambda_i^*, \quad i = \nu+1, \dots, 2n+m$$
where $\lambda_i^*$, $i=\nu+1,\dots,n$, are the zeros of $P(s)$, and $\lambda_i^*$, $n+1,\dots,2n+m$, are the poles of the nominal closed-loop with $P_n(s)$ and $C(s)$.

This argument shows that, if there is no $\lambda_i^*$ on the imaginary axis, the conditions (a), (b), and (c) are also necessary for robust stability with sufficiently small $\tau$.
It is also seen by \eqref{eq:lambda} that, when $p_f(s)$ is Hurwitz, the poles $\lambda_i(\tau)$, $i=1,\dots,\nu$, escapes to the negative infinity as $\tau \to 0$.
Therefore, with large bandwidth of $Q(s)$, the closed-loop system shows two-time scale behavior.
In fact, it turns out that $p_f(s)$ is the characteristic polynomial of the fast dynamics called `boundary-layer system' in the singular perturbation analysis with $\tau$ being the parameter of singular perturbation.

\subsection{Design of $Q(s)$ for robust stability}

In order to satisfy the condition (c) of Theorem \ref{thm:1}, one can choose $\{a_i : i=1, \cdots, \nu-1\}$ such that
$$s^{\nu-1} + a_{\nu-1} s^{\nu-2} + \cdots + a_2 s + a_1$$
be a Hurwitz polynomial, and then pick $a_0 > 0$ sufficiently small.
Then, the polynomial $p_f(s)$ remains Hurwitz for all variation of $g \in [\underline g,\overline g]$.

This can be justified by, for example, the circle criterion.
With $G(s) := a_0/(s^\nu + a_{\nu-1} s^{\nu-1} + \cdots + a_1 s)$ and a static gain $(g/g^*)$ that belongs to the sector $[\underline g/g^*,\overline g/g^*]$, the characteristic polynomial of the closed-loop system becomes $s^\nu + a_{\nu-1} s^{\nu-1} + \cdots + a_1 s + (g/g^*)a_0$.
Therefore, if the Nyquist plot of $G(s)$ does not enter nor encircle the closed disk in the complex plane whose diameter is the line segment connecting $-g^*/\underline g$ and $-g^*/\overline g$,
then $p_f(s)$ is Hurwitz for all variation of $g \in [\underline g,\overline g]$.
Since $G(s)$ has one pole at the origin and the rest poles have negative real parts, its Nyquist plot is bounded to the direction of real axis.
Therefore, by choosing $a_0$ sufficiently small, the Nyquist plot is disjoint from and does not encircle the disk.

\section{Disturbance Observer for Nonlinear System}

\subsection{Intuitive introduction of the DOB for nonlinear system}

DOB for nonlinear systems inherits all the ingredients and properties of the DOB for linear systems.
The DOB can be constructed for a class of systems that can be represented by the Byrnes-Isidori normal form in certain coordinates as
\begin{equation*}
P: \quad \left\{ \quad
\begin{aligned}
y = x_1, \qquad \qquad \dot x_i &= x_{i+1}, \qquad i = 1, \cdots, \nu-1, \\
\dot x_\nu &= f(x,z) + g(x,z) (u + d), \\
\dot z &= h(x,z,d_z) 
\end{aligned} \right.
\end{equation*}
where $u \in \R$ is the input, $y \in \R$ is the measured output, $x = [x_1,\cdots,x_\nu]^T \in \R^\nu$ and $z \in \R^{n-\nu}$ are the states of the $n$-th order system, and unknown external disturbances are denoted by $d$ and $d_z$.
The disturbances and their first time-derivatives are assumed to be bounded.
The functions $f$ and $g$, and the vector field $h$ contain uncertainty.
The corresponding nominal model of the plant is considered as
\begin{equation*}
P_n: \quad \left\{ \quad
\begin{aligned}
y = x_1, \qquad \qquad \dot x_i &= x_{i+1}, \qquad i = 1, \cdots, \nu-1, \\
\dot x_\nu &= f_n(x,z) + g_n(x,z) \bar u \\
\dot z &= h_n(x,z,0)
\end{aligned} \right.
\end{equation*}
where $\bar u$ represents the input to the nominal model, and the nominal $f_n$, $g_n$, and $h_n$ are known.
Let us assume all functions and vector fields are smooth.

\begin{assumption}\label{ass:1}
\begin{enumerate}
	\item [(a)] A baseline feedback controller 
	\begin{equation*}
C: \quad \left\{ \quad
\begin{aligned}
\dot \eta &= \Pi(\eta,y) \quad \in \R^m, \\
\bar u &= \pi(\eta,y)
\end{aligned} \right.
\end{equation*}
stabilizes the nominal model $P_n$.
	\item [(b)] The zero dynamics $\dot z = h(x,z,d_z)$ is input-to-state stable (ISS) from $x$ and $d_z$ to the state $z$.

\end{enumerate}
\end{assumption}

The underlying idea of the DOB is that, if one can apply the input $u_{desired}$ to the plant $P$, which is generated by
\begin{align}\label{eq:desired}
\begin{split}
\dot {\bar z} &= h_n(x,\bar z) \\
u_{desired} &= -d + \frac{1}{g(x,z)} (-f(x,z) + f_n(x,\bar z) + g_n(x,\bar z) \bar u)
\end{split}
\end{align}
then the plant $P$ with \eqref{eq:desired} behaves identical to the nominal model $P_n$ plus $\dot z = h(x,z,d_z)$.
Then, the plant $P$ with \eqref{eq:desired} interacts with the baseline controller $C$, and so, it is stabilized, while the plant's zero dynamics $\dot z = h(x,z,d_z)$ becomes stand-alone.
However, the zero dynamics is ISS, and so, the state $z$ is bounded under the bounded inputs $x$ and $d_z$.

This idea is, however, not implementable because $d$, $f$, and $g$, as well as the states $x$ and $z$ in \eqref{eq:desired}, are unknown.
It turns out that the role of the DOB is to {\em estimate} the state $x$ and the signal $u_{desired}$.
In the linear case, the structure of the DOB in Fig.~\ref{fig:lindob} actually does this job \citep{Shim16}.

\subsection{Implementation of the DOB}

The idea of estimating $x$ and $u_{desired}$ is realized in the semi-global sense.
Suppose that $S_0 \subset \R^{n+m}$ be a compact set for possible initial conditions $(x(0),z(0),\eta(0))$, and $U$ be a compact subset of region of attraction for the nominal closed-loop system $P_n$ and $C$, which is assumed to contain a compact set $S$ that is strictly larger than $S_0$.
Let $S_z$ and $U_x$ be the projections of $S$ and $U$ to the $z$-axis and the $x$-axis, respectively, and let $M_d$ and $M_{d_z}$ be an upper bounds for the norms of $d$ and $d_z$, respectively.
Uncertain $f$, $g$, and $h$ are supposed to belong to the sets ${\mathcal F}$, ${\mathcal G}$, and ${\mathcal H}$, respectively, which are defined as follows.
Let ${\mathcal H}$ be a collection of uncertain $h$.
Then there is a compact set $Z_h \subset \R^{n-\nu}$ to which the state $z(t)$ belongs, where $z(t)$ is the solution to each ISS system $\dot z = h(u_1, z, u_2)$, $h \in {\mathcal H}$, with any $z(0) \in S_z$ and any bounded inputs $u_1(t) \in U_x$ and $\|u_2(t)\| \le M_{d_z}$.
Assume that ${\mathcal H}$ is such that there is a compact set $Z \supset \cup_{h \in {\mathcal H}} Z_h$.
The set ${\mathcal F}$ is a collection of uncertain $f$ such that, for every $f \in {\mathcal F}$, there are uniform bounds $M_f$ and $M_{df}$ such that $|f(x,z)| \le M_f$ and $\|\partial f(x,z)/\partial (x,z)\| \le M_{df}$ on $U_x \times Z$.
The set ${\mathcal G}$ is a collection of uncertain $g$ such that, for every $g \in {\mathcal G}$, there are uniform bounds $\underline g$, $\overline g$, and $M_{dg}$ such that $\|\partial g(x,z)/\partial (x,z)\| \le M_{dg}$ and 
$$0 < \underline g \le g(x,z) \le \overline g, \qquad \forall (x,z) \in U_x \times Z.$$

\begin{figure}[h]
	\centering
	\includegraphics[height=5cm]{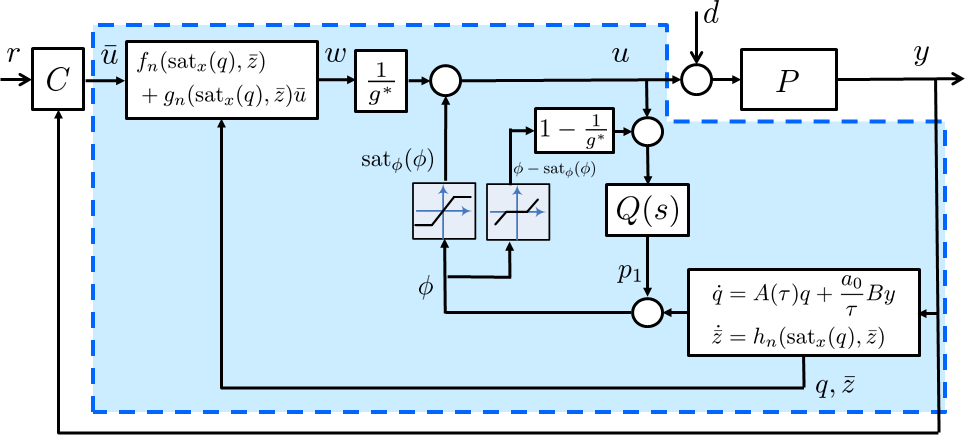}
	\caption{Disturbance observer for a nonlinear plant. The state of $Q(s)$ is $p$.}
	\label{fig:nonlindob}    
\end{figure}

The DOB is illustrated in Fig.~\ref{fig:nonlindob} and is given by
\begin{equation*}
D: \quad \left\{ \quad 
\begin{aligned}
\dot {\bar z} &= h_n({\rm sat}_x(q),\bar z) &  & \in \R^{n-\nu} \\
\dot q &= A(\tau) q + \frac{a_0}{\tau} B y & & \in \R^\nu \\
\dot p &= A(\tau) p + \frac{a_0}{\tau} B \left( \phi - \frac1{g^*}(\phi -{\rm sat}_\phi(\phi)) + \frac1{g^*}w \right) & & \in \R^\nu \\
u &= {\rm sat}_\phi(\phi) + \frac1{g^*}w & &
\end{aligned} \right.
\end{equation*}
where
\begin{align*}
\phi &= p_1 + \frac1{g^*} \left[ \frac{a_0}{\tau^\nu}, \frac{a_1}{\tau^{\nu-1}}, \cdots, \frac{a_{\nu-1}}{\tau} \right] q - \frac1{g^*} \frac{a_0}{\tau^\nu} y \\
w &= f_n({\rm sat}_x(q),\bar z) + g_n({\rm sat}_x(q),\bar z) \bar u \\
A(\tau) &= \begin{bmatrix} 0 & 1 & \cdots & 0 \\
\vdots & \vdots & \ddots & \vdots \\
0 & 0 & \cdots & 1 \\
\frac{-a_0}{\tau^\nu} & \frac{-a_1}{\tau^{\nu-1}} & \cdots & \frac{-a_{\nu-1}}{\tau} \end{bmatrix}, \qquad
B = \begin{bmatrix} 0 \\ \vdots \\ 0 \\ 1 \end{bmatrix}
\end{align*}
and $g^*$ is any constant between $\underline g$ and $\overline g$, $\bar u$ is the output of the baseline controller $C$, and two saturations are globally bounded, continuously differentiable functions such that ${\rm sat}_x(x) = x$ for all $x \in U_x$ and ${\rm sat}_\phi(\phi) = \phi$ for all $\phi \in S_\phi$ where the interval $S_\phi$ is given by
\begin{multline*}
S_\phi = \bigg\{ \left(\frac1{g(z,x)} - \frac1{g^*}\right) (f_n(x,\bar z) + g_n(x,\bar z) \pi(\eta,x_1)) - \frac{f(x,z)}{g(x,z)} - d \\
: z \in Z, (\bar z,x,\eta) \in U, |d| \le M_d, f \in {\mathcal F}, g \in {\mathcal G} \bigg\} .
\end{multline*}
Suppose that $(\bar z(0), q(0), p(0)) \in S_{0z} \times S_{qp}$ where $S_{0z}$ is the projection of $S_0$ to the $z$-axis, and $S_{qp}$ is any compact set in $\R^{2\nu}$.

In practice, choosing the sets like $U$, $Z$, and $S_\phi$ may not be easy.
If so, conservative choice of them works.
Based on repeated simulations or numerical computations, one can take sufficiently large compact sets for them.

If everything is linear, then the above controller $D$ becomes effectively the same as the state-space realization of the linear DOB in Fig.~\ref{fig:lindob}.

\subsection{Robust stability}

\addtocounter{assumption}{-1}
\begin{assumption}
\begin{enumerate}
\item [(c)] The coefficients $a_i$ in the DOB are chosen such that $s^{\nu-1} + a_{\nu-1} s^{\nu-2} + \cdots + a_1$ is Hurwitz and $a_0>0$ is sufficiently small such that the Nyquist plot of $G(s) := a_0/(s^\nu + a_{\nu-1} s^{\nu-1} + \cdots + a_1 s)$ is disjoint from and does not encircle the closed disk in the complex plane whose diameter is the line segment connecting $-g^*/\underline g$ and $-g^*/\overline g$.
\end{enumerate}
\end{assumption}

Under this assumption, the polynomial
$$p_f(s) := s^\nu + a_{\nu-1} s^{\nu-1} + \cdots + a_1 s + \frac{g(x,z)}{g^*} a_0$$
is Hurwitz for all $(x,z) \in U_x \times Z$ and for all uncertain functions $g \in {\mathcal G}$, as discussed in the section `Design of $Q(s)$ for robust stability.'

\begin{thm}\label{thm:nonlin}
Suppose that the conditions (a), (b), and (c) of Assumption \ref{ass:1} hold.
Then, there exists $\tau^*>0$ such that, for all $\tau \le \tau^*$, the closed-loop system of $P$, $C$, and $D$ with $d \equiv 0$ and $d_z \equiv 0$ is stable for all $f \in {\mathcal F}$, $g \in {\mathcal G}$, and $h \in {\mathcal H}$, and the region of attraction for $(x,z,\eta,\bar z,q,p)$ includes $S_0 \times S_{0z} \times S_{qp}$.
\end{thm}

To prove the theorem, one can convert the closed-loop system into the standard singular perturbation form with $\tau$ being the singular perturbation parameter.
Then, it can be seen that the quasi-steady-state system on the slow manifold is simply the nominal closed-loop system $P_n$ and $C$ without external disturbance $d$ plus the actual zero dynamics $\dot z = h(x,z,d_z)$.
Since the quasi-steady-state system is assumed to be stable by Assumption \ref{ass:1}, the overall system is stable with sufficiently small $\tau$ if the boundary-layer system is also stable.
Then, it turns out that the boundary-layer system is linear since all the slow variables such as $x(t)$ and $z(t)$ are treated as frozen parameters, and the characteristic polynomial of the linear boundary-layer system is $p_f(s)$.
For details, see \citep{BS08}.

It can be also seen that, on the slow manifold, all the saturation functions in the DOB become inactive.
The role of the saturations is to prevent the peaking phenomenon \citep{SK91} from transferring into the plant.
Without saturations, the region of attraction may shrink in general as $\tau$ gets smaller, as in \citep{KM86}, and only local stability is achieved.
On the other hand, even if the plant is protected from the peaking components by the saturation functions, some internal components must peak for fast convergence of the DOB states.
In this regard, the role of the dead-zone nonlinearity in Fig.~\ref{fig:nonlindob} is to allow peaking inside the DOB structure.

\subsection{Robust Transient Response}

Additional benefit of the DOB with saturation functions is robustness of transient response.
If the baseline controller $C$ is designed for the nominal model $P_n$ to achieve desired transients such as small overshoot or fast settling time for example, then, similar transients can be obtained for the actual plant $P$ under external disturbances by 
adding the nonlinear DOB to the baseline controller.
This holds true also for linear plants.

\begin{thm}[\citet{BS08}]\label{thm:3}
Suppose that the conditions (a), (b), and (c) of Theorem \ref{thm:nonlin} hold.
For any given $\epsilon > 0$, there exists $\tau^*>0$ such that, for each $\tau \le \tau^*$, the solution of the closed-loop system denoted by $(z(t),\bar z(t), x(t), \eta(t))$, initiated in $S_z \times S$, is bounded and satisfies
$$\left\| \begin{bmatrix} \bar z(t) \\ x(t) \\ \eta(t) \end{bmatrix} - \begin{bmatrix} \bar z_N(t) \\ x_N(t) \\ \eta_N(t) \end{bmatrix} \right\| \le \epsilon, \qquad \forall t \ge 0$$
where the reference $(\bar z_N(t), x_N(t), \eta_N(t))$ is the solution of the nominal closed-loop system of $P_n$ and $C$ with $(\bar z_N(0), x_N(0), \eta_N(0)) = (\bar z(0), x(0), \eta(0))$.
\end{thm}

Since $y = x_1$, this theorem ensures robust transient response that $y(t)$ remains close to its nominal counterpart $y_N(t)$ from the initial time $t=0$.
However, nothing can be said regarding the state $z(t)$ except that it is bounded by the ISS property of the zero dynamics.

Theorem \ref{thm:3} is basically an application of Tikhonov's theorem \citep{H66}.

\section{Discussions}

\begin{itemize}

\item 
The baseline controller $C$ may depend on a reference $r$, which is the case for tracking control.
Theorems \ref{thm:nonlin} and \ref{thm:3} also hold for this case \citep{BS08}.

\item 
If uncertainties are small in modeling so that $f \approx f_n$, $g \approx g_n$, and $h \approx h_n$, then the DOB can be used to estimate the input disturbance $d$.
This is clearly seen from \eqref{eq:desired}.

\item Regarding a design of DOB for multi-input-multi-output plants, refer to \citep{BS09}, which requires well-defined vector relative degree of the plant. 
For the case of extended state observer (ESO), this requirement has been relaxed in \citep{WILK19}.

\item If the external disturbance is a sum of a modeled disturbance, that is generated by a known model, and a unmodeled disturbance that is slowly varying, then the modeled disturbance can be exactly canceled by embedding the known model into the DOB structure while the unmodeled disturbance can still be canceled approximately. This is done by utilizing the internal model principle, and the details can be found in \citep{JPBS14}.

\item For linear systems with mismatched disturbances:
\begin{align*}
\dot {\sf x} &= {\sf A} {\sf x} + {\sf b} u + {\sf E} {\sf d}, & & {\sf x} \in \R^n, \quad u \in \R, \\
y &= {\sf c} {\sf x}, & & {\sf d} \in \R^q, \quad y \in \R,
\end{align*}
one can transfer the disturbances into the input stage by redefining the state combined with the disturbances, if the disturbance ${\sf d}$ is smooth in $t$.
Indeed, there is a coordinate change $[x^T, z^T]^T = \Phi {\sf x} + \Theta [{\sf d}^T, \dot {\sf d}^T, \cdots, ({\sf d}^{(\nu-2)})^T]^T$ from ${\sf x}$ to $(x,z)$ with a nonsingular matrix $\Phi$, and the system becomes
\begin{align*}
y = x_1, \qquad \qquad \dot x_i &= x_{i+1}, \qquad i = 1, \cdots, \nu-1, \\
\dot x_\nu &= F_x x + F_z z + g u + g d \\
\dot z &= H_x x + H_z z + d_z
\end{align*}
where $d$ and $d_z$ are linear combinations of ${\sf d}$ and its derivatives \citep{Shim16}.

\item 
Robust control based on the linear DOB is relatively simple to construct, which is one of the reasons why it is frequently used in industry.
Stability condition is often ignored in practice, but as seen in Theorem \ref{thm:1}, all three conditions are often automatically met.
In particular, with small amount of uncertainty, the condition (c) tends to hold with $g \approx g^*$ for any stable Q-filter because of structural robustness of Hurwitz polynomials.

\item In the case of linear DOB, there is another robust stability condition based on the small-gain theorem \citep{Choi03}.

\item Basic philosophy of DOB is to treat plant's uncertainties and external disturbances together as a lumped disturbance, and to estimate and compensate it. 
This philosophy is shared with other similar approaches such as {\em extended state observer (ESO)} \citep{FK08} and {\em active disturbance rejection control (ADRC)} \citep{H09}.
The DOB also has been reviewed with comparison to other similar methods in \citep{CYGS15}.

\end{itemize}

\section{Summary and Future Directions}

Underlying theory for the disturbance observer is presented.
The analysis is mainly based on large bandwidth of Q-filter.
However, there are cases when the bandwidth cannot be increased in practice because of limited sampling rate in discrete-time implementation.
Further study is necessary to achieve satisfactory performance for discrete-time design of DOB.

\section{Cross References}


\end{document}